\documentclass[aps,reprint,superscriptaddress]{revtex4}
\usepackage{amssymb}
\usepackage{amsmath}

\begin{document}
\baselineskip18pt
\title{Entanglement enhanced distinguishability of coherence-breaking channels}
\author{Long-Mei Yang}
\affiliation{School of Mathematical Sciences,  Capital Normal University,  Beijing 100048,  China}
\author{Tao Li}
\affiliation{School of Science, Beijing Technology and Business University, Beijing 100048, China}
\author{Shao-Ming Fei}
\affiliation{School of Mathematical Sciences,  Capital Normal University,  Beijing 100048,  China}
\affiliation{Max Planck Institute for Mathematics in the Sciences, Leipzig 04103, Germany}
\author{Zhi-Xi Wang}
\affiliation{School of Mathematical Sciences,  Capital Normal University,  Beijing 100048,  China}

\begin{abstract}
\baselineskip18pt
Originated from the superposition principle in quantum mechanics, coherence
has been extensively studied as a kind important resource in quantum information processing.
We investigate the distinguishability of coherence-breaking channels
with the help of quantum entanglement. By explicitly computing the minimal error probability of
channel discrimination, it is shown that entanglement can enhance the capacity of coherence-breaking channel distinguishability with same types for some cases while cannot enhanced for some other cases.
For coherence-breaking channels with different types, the channel distinguishability cannot be enhanced via entanglement.
\end{abstract}

\keywords{coherence-breaking channels; channel distinguishability; entanglement.}

\maketitle
\baselineskip20pt

\noindent{\sf Introduction}~ Quantum coherence is a fundamental aspect of quantum physics,
which encapsulates the defining features of the theory from the superposition principle to quantum correlations \cite{Leggett}.
Quantum coherence is also an essential ingredient in quantum information processing \cite{QCQI}. It constitutes a powerful resource
for quantum metrology \cite{limit,PRL113} and entanglement creation \cite{light,PRL115}, and plays a central role in the emergent fields
such as nanoscale thermodynamics \cite{berg,limitation,PRX5,PRB84} and quantum biology \cite{JPCB,CP54}.
The quantification and implication of quantum coherence have been extensively studied recently \cite{bukf,zhuhj}.
A related important problem is the transformation of quantum coherence under quantum channels.
It is of particular significance to study the class of coherence-breaking channels¡ªtrace preserving
completely positive maps for which the output state is always incoherence \cite{CBC}.
More precisely, a quantum channel $\Phi$ is called coherence breaking if $\Phi(\rho)$ is always incoherent for any density matrix $\rho$.
And one of the key problems in quantum information-processing task is the channel discrimination.
In \cite{optimal, EBC,chi,matt,zhao,cai,li} the optimal discrimination of quantum operations has been investigated.
It has been shown that the distinguishability of entanglement-breaking channels can be enhanced.
Comparing coherence with entanglement, one may naturally ask whether the entanglement can enhance
the distinguishability of the coherence-breaking channel.
In this paper, we study the relation between entanglement and coherence-breaking channels,
and show that entanglement can indeed enhance the distinguishability of coherence-breaking channels with same types for some cases.
For coherence-breaking channels with different types, entanglement cannot better enhance the channel distinguishability.

\noindent{\sf Preliminaries}~
We first recapitulate some concepts required in presenting our main results.
Let $\mathcal{H}$ denote a discrete finite-dimensional complex vector space associated with a quantum system.
A completely positive and trace-preserving (CPTP) map $\Phi$ on a state $\rho$ can be expressed as
$\Phi(\rho)=\sum\limits_nK_n\rho K_n^{\dag}$, where $K_n$ are Kraus operators on $\mathcal{H}$
satisfying $\sum\limits_nK_n^{\dag}K_n=I$ \cite{Kraus}.
A qubit coherence-breaking channel is a CPTP map for which the Kraus operators $K_n$ can be only one of the following three types \cite{yang18}:
\begin{equation}\label{CBC1}
E_{11}=\left(
  \begin{array}{cc}
    1 & 0 \\
    0 & 0 \\
  \end{array}
\right), \ \ \
E_{12}=\left(
  \begin{array}{cc}
    0 & 1 \\
    0 & 0 \\
  \end{array}
\right),
\end{equation}
\begin{equation}\label{CBC2}
E_{21}=\left(
  \begin{array}{cc}
    0 & 0 \\
    0 & e^{i\xi} \\
  \end{array}
\right), \ \ \
E_{22}=\left(
  \begin{array}{cc}
    0 & 0 \\
    e^{i\xi} & 0 \\
  \end{array}
\right),
\end{equation}
or
\begin{equation}\label{CBC3}
E_{31}=\left(
  \begin{array}{cc}
    0 & 0 \\
    -\sin\phi & e^{i\xi}\cos\phi \\
  \end{array}
\right), \ \ \
E_{32}=\left(
  \begin{array}{cc}
    \cos\phi & e^{i\xi}\sin\phi \\
    0 & 0 \\
  \end{array}
\right).
\end{equation}

Given two quantum channels $\Phi_1$ and $\Phi_2$ with a prior probability $p_1$ and $p_2=1-p_1$, respectively.
The problem of optimal discrimination of $\Phi_1$ and $\Phi_2$
can be reformulated into the problem of finding out a state $\rho$ such that the error
probability in the discrimination of the output states $\Phi_1(\rho)$ and $\Phi_2(\rho)$ is minimal \cite{optimal}.
The minimal error probability is given by
\begin{equation}\label{pE1}
p_E^{\prime}=\frac{1}{2}(1-\max\limits_{\rho\in \mathcal{H}}\|p_1\Phi_1(\rho)-p_2\Phi_2(\rho)\|_1),
\end{equation}
where $\|\cdot\|_1$ denotes the trace norm, if no entanglement is taken into consideration.
 However, if we take in account of entanglement, i.e.,
 the system is entangled with another system $\mathcal{K}$, then the minimal error probability is changed to be
\begin{equation}\label{pE2}
p_E=\frac{1}{2}(1-\max\limits_{\rho\in\mathcal{H}\otimes\mathcal{K}}\|p_1(\Phi_1\otimes\mathcal{I})(\rho)-p_2(\Phi_2\otimes\mathcal{I})(\rho)\|_1).
\end{equation}
Here the maximum in \eqref{pE1} and \eqref{pE2} are both achieved by pure states.

Denote $|A\rangle\rangle=\sum\limits_{mn}\langle n|A|m\rangle|n\rangle\otimes|m\rangle=A\otimes I|I\rangle\rangle=I\otimes A^{t}|I\rangle\rangle$ \cite{optimal}.
Define a pure bipartite state $\rho=|\zeta\rangle\rangle\langle\langle\zeta|$ with ${\rm Tr}\rho={\rm Tr}(\zeta^{\dag}\zeta)=1$,
then $p_E$ defined in Eq.\eqref{pE2} can be rewritten as
\begin{equation}\label{pE4}
p_E=\frac{1}{2}(1-\max\limits_{{\rm Tr}[\zeta^{\dag}\zeta]}\|I\otimes\zeta^{\dag}\triangle I\otimes\zeta^*\|_1),
\end{equation}
where $\bigtriangleup=p_1\sum\limits_n|K_n^{(1)}\rangle\rangle\langle\langle K_n^{(1)}|-p_2\sum\limits_m |K_m^{(2)}\rangle\rangle\langle\langle K_m^{(2)}|$
, and $\{K_n^{(1)}\}$ and $\{K_m^{(2)}\}$ are the Kraus operators of quantum operations, respectively.
Also, $p_E\leqslant\frac{1}{2}(1-\frac{1}{d}\|\Delta\|_1)$ with $d=\dim(\mathcal{H})$, where equality holds for quantum operations with the form of
$\Phi_i(\rho)=\sum\limits_n q_n^{(i)}U_n\rho U_n^{\dag}$, and $\sum\limits_n q_n^{(i)}=1$.
Here, $\{U_n\}$ is a set of orthogonal unitaries with ${\rm Tr}(U_m^{\dag}U_n)=d\delta_{n,m}$.
In this case, $\Delta=\sum\limits_n r_n|U_n\rangle\rangle\langle\langle U_n|$ with $r_n=p_1q_n^{(1)}-p_2q_n^{(2)}$.
Then $p_E$ can also be written in the form
\begin{equation}\label{pE3}
p_E=\frac{1}{2}(1-\max\limits_{P\geqslant0, \ {\rm Tr}[P^2]=1}\|I\otimes P\bigtriangleup I\otimes P\|_1),
\end{equation}
where for qubit channels, $P$ can be written as
$P=\left(
           \begin{array}{cc}
             x & z \\
             z^* & y \\
           \end{array}
         \right)$
with $x,y\geqslant0$, $xy\geqslant|z|^2$ and $x^2+y^2+2|z|^2=1$.
Particularly, one has $x+y=1$ and $|z|=\sqrt{xy}$ if ${\rm rank} (P)=1$.
Also, one finds that the entanglement is not needed for optimal discrimination if and only if the maximum in \eqref{pE4}
can be achieved by an operator $P$ with ${\rm rank} (P)=1$.

\noindent{\sf Coherence-breaking channels of the same type}~
We first consider qubit coherence-breaking channels of the the same types given in \eqref{CBC1}, \eqref{CBC2} and \eqref{CBC3}.
Obviously the discrimination for two channels of the form either \eqref{CBC1} or \eqref{CBC2} is trivial.
We only need to study the problem for two coherence-breaking channels that both are of the form \eqref{CBC3}.
Let $\Phi_i$, $i=1,2$, be two coherence-breaking channels with, respectively,
$E_1^{(i)}=\left(
  \begin{array}{cc}
    0 & 0 \\
    -\sin\phi_i & e^{i\xi_i}\cos\phi_i \\
  \end{array}
\right), \ \ \
E_2^{(i)}=\left(
  \begin{array}{cc}
    \cos\phi_i & e^{i\xi_i}\sin\phi_i \\
    0 & 0 \\
  \end{array}
\right)$ for $i=1,2$.
Then,
\begin{equation}
\begin{array}{rl}
\bigtriangleup &=p_1(|E_1^{(1)}\rangle\rangle\langle\langle E_1^{(1)}|+|E_2^{(1)}\rangle\rangle\langle\langle E_2^{(1)}|)-
p_2(|E_1^{(2)}\rangle\rangle\langle\langle E_1^{(2)}|+|E_2^{(2)}\rangle\rangle\langle\langle E_2^{(2)}|)\\[3mm]
&=\left(
   \begin{array}{cccc}
     \bigtriangleup_{11} & \bigtriangleup_{12} & 0 & 0 \\
     \bigtriangleup_{12}^{*} & \bigtriangleup_{22} & 0 & 0 \\
     0 & 0 & \bigtriangleup_{22} & -\bigtriangleup_{12} \\
     0 & 0 & -\bigtriangleup_{12}^{*} & \bigtriangleup_{11} \\
   \end{array}
 \right),
 \end{array}
\end{equation}
where $\bigtriangleup_{11}=p_1\cos^2\phi_1-p_2\cos^2\phi_2$, $\bigtriangleup_{12}=p_1e^{-i\xi_1}\sin\phi_1\cos\phi_1-p_2e^{-i\xi_2}\sin\phi_2\cos\phi_2$
and $\bigtriangleup_{22}=p_1\sin^2\phi_1-p_2\sin^2\phi_2$.
The singular values of $\bigtriangleup$ are
$$
s_0(\bigtriangleup)=s_1(\bigtriangleup)=
\frac{1}{2}\left|p_1-p_2+M~\right|,
\ \ \
s_2(\bigtriangleup)=s_3(\bigtriangleup)=
\frac{1}{2}\left|p_1-p_2-M~\right|,
$$
where $$M=\sqrt{p_1^2+p_2^2-2p_1p_2\cos2\phi_1\cos2\phi_2-p_1p_2(e^{i\xi_1-i\xi_2}+e^{i\xi_2-i\xi_1})\sin2\phi_1\sin2\phi_2}.$$
Thus,
\begin{equation}
\sum\limits_{i=0}^3s_i(\bigtriangleup)=2\max\left\{|p_1-p_2|,
M~\right\}.
\end{equation}
Then
\begin{equation}
\sum\limits_{i=0}^3s_i(\bigtriangleup)=2\sqrt{p_1^2+p_2^2-2p_1p_2\cos2\phi_1\cos2\phi_2-2p_1p_2\cos(\xi_1-\xi_2)\sin2\phi_1\sin2\phi_2},
\end{equation}
since
\begin{equation}
\begin{array}{rl}
& \ \ \ p_1^2+p_2^2-2p_1p_2\cos2\phi_1\cos2\phi_2-p_1p_2(e^{i\xi_1-i\xi_2}+e^{i\xi_2-i\xi_1})\sin2\phi_1\sin2\phi_2\\[2.0mm]
&=p_1^2+p_2^2-2p_1p_2\cos2\phi_1\cos2\phi_2-2p_1p_2\cos(\xi_1-\xi_2)\sin2\phi_1\sin2\phi_2\\[2.0mm]
&\geqslant\min\{p_1^2+p_2^2-2p_1p_2\cos2(\phi_1-\phi_2), \ p_1^2+p_2^2-2p_1p_2\cos2(\phi_1+\phi_2)\}\\[2.0mm]
&\geqslant(p_1-p_2)^2.
\end{array}
\end{equation}

Note that for a matrix $A$, $\|A\|_1={\rm Tr}\sqrt{A^{\dag}A}=\max\limits_{U}|{\rm Tr}[UA]|=\sum\limits_is_i(A)$,
where $U$ is a unitary operator and $\{s_i(A)\}$ denote the singular values of $A$.
For the strategy without entangled input, we get
$p_E^{\prime}=\displaystyle\frac{1}{2}(1-\max\limits_{P^{\prime}}\|I\otimes P^{\prime}\bigtriangleup I\otimes P^{\prime}\|_1)$
with $P^{\prime}=\left(
                   \begin{array}{cc}
                     x & \sqrt{x(1-x)}e^{i\phi} \\
                     \sqrt{x(1-x)}e^{-i\phi} & 1-x \\
                   \end{array}
                 \right)$,
$0\leqslant x\leqslant 1$ and $0\leqslant\phi\leqslant 2\pi$.
The singular values of $\bigtriangleup^{\prime}=I\otimes P^{\prime}\bigtriangleup I\otimes P^{\prime}$ are given by
\begin{equation*}
\begin{array}{rl}
 s_0(\bigtriangleup^{\prime})
= &\left|(p_1-p_2)-(p_1\cos^2\phi_1-p_2\cos^2\phi_2)x +(p_1\sin^2\phi_1-p_2\sin^2\phi_2)(1-x)\right.\\
&\hskip6cm \left. +(re^{i\phi}+r^*e^{-i\phi})\sqrt{x(1-x)}\right|,
\end{array}
\end{equation*}

\begin{equation*}
\begin{array}{rl}
s_1(\bigtriangleup^{\prime})=
& \left|(p_1\cos^\phi_1-p_2\cos^2\phi_2)x  +(p_1\sin^2\phi_1-p_2\sin^2\phi_2)(1-x)\right.\\
&\hskip6cm \left. +(re^{i\phi}+r^*e^{-i\phi})\sqrt{x(1-x)}~\right|,
\end{array}
\end{equation*}
and $s_2(\bigtriangleup^{\prime})=s_3(\bigtriangleup^{\prime})=0$ with
$r=p_1e^{i\xi_1}\sin\phi_1\cos\phi_1-p_2\sin\phi_2\cos\phi_2$. Therefore,
\begin{equation}
\begin{array}{rl}
&\sum\limits_{i=0}^3s_i(\bigtriangleup^{\prime})=\max\left\{|p_1-p_2|, \sqrt{(re^{i\phi}+r^*e^{-i\phi})^2+(p_1\cos2\phi_1-p_2\cos2\phi_2)^2}|\right\}\\[2.0mm]
&\ \ \ \ \ \ \ \ \ \ \ \ \ \ =p_1^2+p_2^2-2p_1p_2\cos2\phi_1\cos2\phi_2+2p_1p_2\sin2\phi_1\sin2\phi\cos(\xi_1-\xi_2).
\end{array}
\end{equation}
Although the necessary and sufficient condition for $p_E<p_E^{\prime}$ can be difficult to
calculate explicitly in general, we can find that $p_E\leqslant\displaystyle\frac{1}{2}(1-\frac{1}{2}\sum\limits_{i=0}^3s_i(\bigtriangleup))
<p_E^{\prime}=\displaystyle\frac{1}{2}(1-\sum\limits_{i=0}^3s_i(\bigtriangleup^{\prime}))$ for
$\sin2\phi_1\sin2\phi_2\cos(\phi_1-\phi_2)<0$.
Hence it is clear that the entanglement can indeed enhance the distinguishability of the coherence-breaking channels \eqref{CBC3}.

\textit{Example}.
Consider the case that the prior probability $p_1=p_2=\displaystyle\frac{1}{2}$, and $\Phi_1$ and $\Phi_2$
satisfy conditions $\sin2\phi_1=-\sin2\phi_2=\displaystyle\frac{\sqrt{2}}{2}$, $\cos2\phi_1=\cos2\phi_2=\displaystyle\frac{\sqrt{2}}{2}$,
and $\xi_1=\xi_2=0$.
We have, $\sum\limits_{i=0}^3s_i(\bigtriangleup^{\prime})=0$ and $\sum\limits_{i=0}^3s_i(\bigtriangleup)=1$.
Then, $p_E\leqslant\displaystyle\frac{1}{4}<p_E^{\prime}=\displaystyle\frac{1}{2}$.

\noindent{\sf Coherence-breaking channels of different types}~
We study now coherence-breaking channels with different forms. We first assume $\Phi_1$ and $\Phi_2$  are defined by \eqref{CBC1} and \eqref{CBC2}, respectively.
In this case, it can be found that
\begin{equation}
\begin{array}{rl}
\bigtriangleup & =p_1(|E_1^{(1)}\rangle\rangle\langle\langle E_1^{(1)}|+|E_2^{(1)}\rangle\rangle\langle\langle E_2^{(1)}|)-
p_2(|E_1^{(2)}\rangle\rangle\langle\langle E_1^{(2)}|+|E_2^{(2)}\rangle\rangle\langle\langle E_2^{(2)}|)\\
& =\left(
   \begin{array}{cccc}
     p_1 & 0 & 0 & 0 \\
     0 & p_1 & 0 & 0 \\
     0 & 0 & -p_2 & 0 \\
     0 & 0 & 0 & -p_2 \\
   \end{array}
 \right).
\end{array}
\end{equation}
The singular values of $\bigtriangleup$ are: $s_0(\bigtriangleup)=s_1(\bigtriangleup)=p_1, \ s_2(\bigtriangleup)=s_3(\bigtriangleup)=p_2$.
Thus, $p_E=\displaystyle\frac{1}{2}(1-\frac{1}{2}\sum\limits_{i=0}^3s_i(\bigtriangleup))=0$.
That is to say, the entanglement can help to distinguish these two channels perfectly.

For the strategy without entanglement, we get
$p_E^{\prime}=\displaystyle\frac{1}{2}(1-\max\limits_{P^{\prime}}\|I\otimes P^{\prime}\bigtriangleup I\otimes P^{\prime}\|_1)$, with $P^{\prime}$ defined as above.
The singular values of $\bigtriangleup^{\prime}=I\otimes P^{\prime}\bigtriangleup I\otimes P^{\prime}$ are given by:
$s_0(\bigtriangleup^{\prime})=p_1$, $s_1(\bigtriangleup^{\prime})=p_2$ and
$s_2(\bigtriangleup^{\prime})=s_3(\bigtriangleup^{\prime})=0$. Hence we have
$p_E^{\prime}=\displaystyle\frac{1}{2}(1-\sum\limits_{i=0}^3s_i(\bigtriangleup^{\prime}))=0$,
which shows that with or without the help of entanglement, the channel can always perfectly distinguished.

Now consider channels $\Phi_1$ and $\Phi_2$ defined by \eqref{CBC1} and \eqref{CBC3}, respectively. We have
\begin{equation}
\begin{array}{rl}
\bigtriangleup &=p_1(|E_1^{(1)}\rangle\rangle\langle\langle E_1^{(1)}|+|E_2^{(1)}\rangle\rangle\langle\langle E_2^{(1)}|)-
p_2(|E_1^{(2)}\rangle\rangle\langle\langle E_1^{(2)}|+|E_2^{(2)}\rangle\rangle\langle\langle E_2^{(2)}|)=\\[6mm]
&\left(
   \begin{array}{cccc}
     p_1-p_2\cos^2\phi_2 & -p_2e^{-i\xi}\sin\phi_2\cos\phi_2 & 0 & 0 \\
     -p_2e^{i\xi}\sin\phi_2\cos\phi_2 & p_1-p_2\sin^2\phi_2 & 0 & 0 \\
     0 & 0 & -p_2\sin^2\phi_2 & p_2e^{-i\xi}\sin\phi_2\cos\phi_2 \\
     0 & 0 & p_2e^{i\xi}\sin\phi_2\cos\phi_2 & -p_2\cos^2\phi_2 \\
   \end{array}
 \right).
\end{array}
\end{equation}
The singular values of $\bigtriangleup$ are: $s_0(\bigtriangleup)=p_1$, $s_1(\bigtriangleup)=p_2$,
$s_2(\bigtriangleup)=|p_1-p_2|$ and $s_3(\bigtriangleup)=0$.
Thus, $\sum\limits_{i=0}^3s_i(\bigtriangleup)=1+|p_1-p_2|$.
While for the case without entanglement, the singular values of $\bigtriangleup^{\prime}=I\otimes P^{\prime}\bigtriangleup I\otimes P^{\prime}$ are given by
$s_0(\bigtriangleup^{\prime})=|(p_1-p_2\cos^2\phi_2)x+(p_1-p_2\sin^2\phi_2)(1-x)-(e^{i\xi+i\phi}+e^{-i\xi-i\phi})p_2\sin\phi_2\cos\phi_2\sqrt{x(1-x)}|$,
$s_1(\bigtriangleup^{\prime})=|(e^{i\xi+i\phi}+e^{-i\xi-i\phi})p_2\sin\phi_2\cos\phi_2\sqrt{x(1-x)}-p_2x\sin^2\phi_2-p_2(1-x)\cos^2\phi_2|$
and $s_2(\bigtriangleup^{\prime})=s_3(\bigtriangleup^{\prime})=0$.
Thus
$\sum\limits_{i=0}^3s_i(\bigtriangleup^{\prime})=\max\left\{\left|p_1-p_2\right|, \
\left|\cos(\xi+\phi)\sin2\phi_2\sqrt{x(1-x)}+x\cos^2\phi_2+(1-x)\sin^2\phi_2-1\right| \right\}=1.$
Namely, the entanglement cannot better enhance the distinguishability of the coherence-breaking channels,
since $p_E^{\prime}=\displaystyle\frac{1}{2}(1-\sum\limits_{i=0}^3s_i(\bigtriangleup^{\prime}))=0$ and $p_E\leqslant\displaystyle\frac{1}{2}(1-\frac{1}{2}\sum\limits_{i=0}^3s_i(\bigtriangleup))=\displaystyle\frac{1}{4}-\displaystyle\frac{1}{4}|p_1-p_2|$.

One may also consider channels $\Phi_1$ and $\Phi_2$ given by \eqref{CBC2} and \eqref{CBC3} respectively, and obtain similar conclusions:
entanglement cannot better enhance the ability of discrimination of two channels of this kind.

\noindent{\sf Conclusion and discussion}~
We have investigated the discrimination of two coherence-breaking channels based on error probabilities.
It has been shown that entanglement cannot better enhance the channel distinguishability of different types while
entanglement can improve the channel discrimination of the coherence-breaking channels of type \eqref{CBC3}.
However, there indeed exist coherence-breaking channels with same type whose distinguishability  cannot be enhanced via entanglement.
Let us consider two Pauli channels $\Phi_i(\rho)=\sum\limits_{\alpha=0}^3q_i^{\alpha}\sigma_{\alpha}\rho\sigma_{\alpha}$
with a prior probability $p_1$ and $p_2=1-p_1$, where $\{\sigma_0,\sigma_1,\sigma_2,\sigma_3\}=\{I,\sigma_x,\sigma_y,\sigma_z\}$
and $\sum\limits_{\alpha=0}^3q_i^{\alpha}=1$.
Sacchi\cite{optimal} shown that entanglement strictly improves the discrimination if and only if $\prod\limits_{i=0}^3r_\alpha<0$,
with  $r_\alpha=p_1q_1^{\alpha}-p_2q_2^{\alpha}$.
Assume $\Phi_1$ and $\Phi_2$ are both coherence-breaking Pauli channels.
We have $q_i^{(0)}+q_i^{(1)}-q_i^{(2)}-q_i^{(3)}=q_i^{(0)}-q_i^{(1)}+q_i^{(2)}-q_i^{(3)}=0$, $i=1,2$.
Thus, $r_0=r_3=p_1q_1^{(0)}-p_2q_2^{(0)}$ and $r_1=r_2=p_1q_1^{(1)}-p_2q_2^{(1)}$.
Hence, $\prod\limits_{i=0}^3r_\alpha\geqslant0$, which implies that the entanglement cannot improve the channel discrimination in this case.

\bigskip
\noindent{\bf Acknowledgments}\, \,  This work is supported by the NSF of China under Grant No.11675113, Research Foundation for
Youth Scholars of Beijing Technology and Business University QNJJ2017-03 and Beijing Municipal Education Commission under Grant Nos.KM201810011009 and KZ201810028042.

\end{document}